\documentclass[12pt]{article}

\usepackage{graphicx}
\usepackage{amsmath}
\usepackage{hhline}
\usepackage{amssymb}   
\usepackage{times}
\usepackage{cite}
\usepackage{array}
\usepackage{multirow} 
\usepackage{color}
\definecolor{rltred}{rgb}{0.75,0,0}
\definecolor{rltgreen}{rgb}{0,0.5,0}
\definecolor{rltblue}{rgb}{0,0,0.75}
\definecolor{lightblue}{rgb}{0.72,0.95,0.96}
\usepackage{colordvi}
\usepackage[english]{babel}

\newif\ifpdf
\ifx\pdfoutput\undefined
    \pdffalse          
\else
    \pdfoutput=1       
    \pdftrue
\fi

\usepackage{lineno}
\ifpdf
\DeclareGraphicsExtensions{.pdf,.png,.mps,.eps}
\usepackage{thumbpdf}
\usepackage[pdftex,
        colorlinks=true,
        urlcolor=rltblue,       
        filecolor=rltgreen,     
        linkcolor=rltred,       
        pdftitle={Muon Pair Production in Electron-Proton Collisions},
        pdfauthor={H1},
        pdfsubject={Muon Pair Production},
        pdfkeywords={High-Energy Physics, Particle Physics},
         pagebackref,
        pdfpagemode=None,
        bookmarksopen=true]{hyperref}

\else
\DeclareGraphicsExtensions{.eps,.pdf,.png,.mps}
\fi

\def\PLB{{\em Phys. Lett.}   {\bf B}}

\newlength{\dinwidth}
\newlength{\dinmargin}
\newcommand{\promille}{%
  \relax\ifmmode\promillezeichen
        \else\leavevmode\(\mathsurround=0pt\promillezeichen\)\fi}
\newcommand{\promillezeichen}{%
  \kern-.05em%
  \raise.5ex\hbox{\the\scriptfont0 0}%
  \kern-.15em/\kern-.15em%
  \lower.25ex\hbox{\the\scriptfont0 00}}

\begin{document}

\begin{titlepage}

\noindent
DESY-03-159 \hspace{10.5 cm} ISSN 0418-9833 \\
October 2003\\

\vspace{2cm}

\begin{center}
\begin{Large}

{\bf  Muon Pair Production \\
 in ep Collisions at HERA}

\vspace{2cm}

H1 Collaboration

\end{Large}
\end{center}

\vspace{2cm}

\begin{abstract}

Cross sections for  the production of two isolated muons up to high di-muon masses
are measured in $ep$ collisions at HERA with the H1 detector in a data sample 
corresponding to an integrated luminosity of 71~pb$^{-1}$ 
at a centre of mass energy of $\sqrt{s}=319$~GeV. 
The results are in good agreement with Standard Model predictions, the  dominant process  being 
photon-photon interactions.
Additional muons or electrons are searched for in events with two high 
transverse momentum muons using the full data sample corresponding 
to 114~pb$^{-1}$, where data at $\sqrt{s}=301$~GeV and $\sqrt{s}=319$~GeV are combined.
Both the di-lepton sample and the tri-lepton sample agree well with the predictions.

\end{abstract}

\vspace{1.5cm}

\begin{center}
To be submitted to \PLB
\end{center}

\end{titlepage}
\begin{flushleft}

A.~Aktas$^{10}$,               
V.~Andreev$^{24}$,             
T.~Anthonis$^{4}$,             
A.~Asmone$^{31}$,              
A.~Babaev$^{23}$,              
S.~Backovic$^{35}$,            
J.~B\"ahr$^{35}$,              
P.~Baranov$^{24}$,             
E.~Barrelet$^{28}$,            
W.~Bartel$^{10}$
% ...

\bigskip{\it
 $ ^{1}$ I. Physikalisches Institut der RWTH, Aachen, Germany$^{ a}$ \\
% ...
$ ^{4}$ Inter-University Institute for High Energies ULB-VUB, Brussels;
          Universiteit Antwerpen (UIA), Antwerpen; Belgium$^{ c}$ \\
% ...
$ ^{10}$ DESY, Hamburg, Germany \\
% ...
$ ^{23}$ Institute for Theoretical and Experimental Physics,
          Moscow, Russia$^{ l}$ \\
 $ ^{24}$ Lebedev Physical Institute, Moscow, Russia$^{ e}$ \\
% ...
$ ^{28}$ LPNHE, Universit\'{e}s Paris VI and VII, IN2P3-CNRS,
          Paris, France \\
% ...
$ ^{31}$ Dipartimento di Fisica Universit\`a di Roma Tre
          and INFN Roma~3, Roma, Italy \\
% ...
$ ^{35}$ DESY, Zeuthen, Germany \\
% ...
$ ^a$ Supported by the Bundesministerium f\"ur Bildung und Forschung, FRG,
      under contract numbers 05 H1 1GUA /1, 05 H1 1PAA /1, 05 H1 1PAB /9,
      05 H1 1PEA /6, 05 H1 1VHA /7 and 05 H1 1VHB /5 \\
% ...
$ ^c$ Supported by FNRS-FWO-Vlaanderen, IISN-IIKW and IWT \\
% ...
$ ^e$ Supported by the Deutsche Forschungsgemeinschaft \\
% ...
$ ^l$ Partially Supported by Russian Foundation
      for Basic Research, grant    no. 00-15-96584 \\
}

\end{flushleft}

\newpage

\section{Introduction}
\noindent
Muon pair production in electron proton scattering proceeds mainly via two-photon interactions, 
\begin{math} \gamma\gamma\rightarrow\mu^+ \mu^- \end{math}, where the 
incoming photons are radiated from the beam particles~\cite{Vermaseren:1983cz}.
It is important to check the quantitative agreement between experiment 
and theory in this process, since the understanding of this source of muons is vital in
any search for anomalous muon production~\cite{Adloff:1998aw,Accomando:1993ar}.
The clean experimental signature and the precise Standard Model  prediction provide 
high sensitivity in such searches. 
In an analysis of multi-electron production~\cite{H1:electrons}, six outstanding events,
three di-electron and three tri-electron events, were observed with di-electron masses above 100~GeV,
a region in which the Standard Model prediction is low. 
A comparison with di-muon production in the same experiment is therefore particularly interesting.

In this paper, a study of muon pair production in electron\footnote{In this paper ``electron'' refers
to both electron and positron, if not otherwise stated.}
proton scattering ($ep\rightarrow e\mu\mu X$) is presented using the H1 detector at the $ep$ collider
HERA.
The main part of this analysis is based on data with an integrated luminosity of 70.9
pb$^{-1}$ collected with an electron energy of 27.6~GeV and a proton energy  $E_p=920$~GeV ($\sqrt{s}=319$~GeV). 
These data were  recorded in the years 1999 and 2000 in  
positron proton scattering (60.8~pb$^{-1}$) and electron proton scattering (10.1~pb$^{-1}$).
Differential cross sections as functions of
the invariant mass of the muon pair $M_{\mu\mu}$, the muon transverse momenta $P_{t}^{\mu}$
and the transverse momentum $P_{t}^{X}$ of the hadronic system $X$
are measured for $M_{\mu\mu}>5$~GeV. 
Results are also given for elastic and inelastic  muon pair production separately.
In addition, events with high di-lepton masses are studied in $\mu\mu$ and $\mu\mu e$ event samples with 
cuts adapted to the multi-electron analysis~\cite{H1:electrons}.
For this analysis the data at $\sqrt{s}=301$~GeV ($E_p=820$~GeV) and $\sqrt{s}=319$~GeV from the years  1994 - 2000 
are combined, yielding a total luminosity of 113.7~pb$^{-1}$.

\section{Standard  Model  Processes}
\label{sec:simulation}

\begin{figure}[h]
  \begin{center}
    \includegraphics[width=13.5cm]{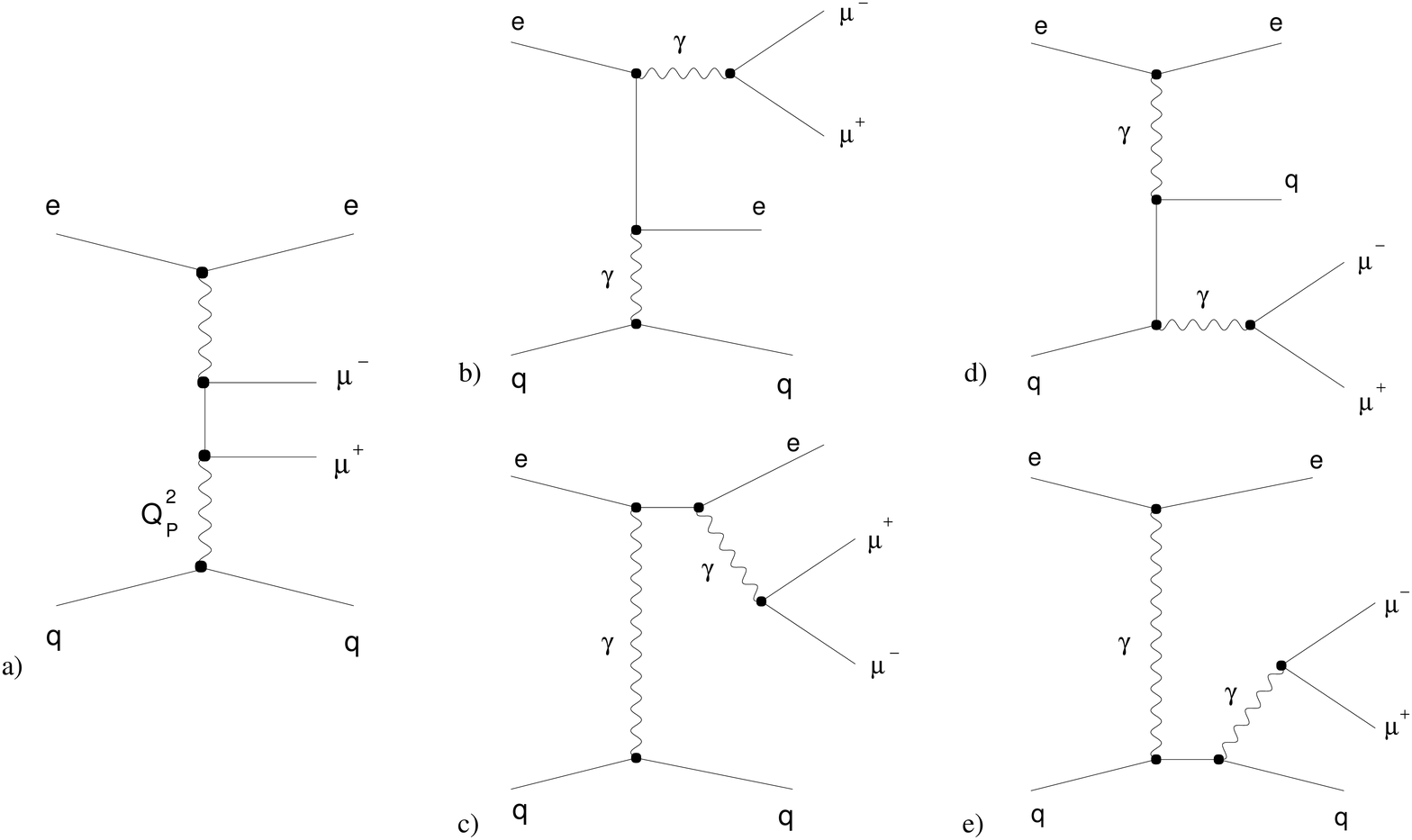}
    \caption{ Feynman diagrams for muon pair production in electromagnetic $eq$
      interactions, shown for the inelastic case where the scattering takes
      place from a single quark, such that the proton dissociates to a system
      $X$. In the elastic case, the proton scatters coherently. In the general
      electroweak case, any of the photon propagators can be replaced by a $Z$ boson propagator. }
    \label{fig:feynman:elweak}
    \label{fig:feynman:bh}
 \end{center}
\end{figure}

The dominant process for the production of muon pairs in $ep$ interactions is
the two-photon reaction illustrated in the Feynman
diagram shown in figure~\ref{fig:feynman:bh}a. 
Due to the photon propagators, the momentum transfer to the scattered
particles is generally small. 
While the calculation of the corresponding processes in $e^{+}e^{-}$ scattering is
rather straightforward, the hadronic structure of the proton must be
taken into account in $ep$ scattering. 
The Feynman diagrams (b)-(e) in figure \ref{fig:feynman:elweak} show additional
sources of muon pairs, which are less important than the two-photon reaction~(a).
They represent the four bremsstrahlung amplitudes in $eq$ 
scattering with conversion of the radiated photon into a muon pair.
The diagrams with the photons radiated from the electron lines can be
viewed as Compton scattering of the electron and
the photon, which is exchanged with the quark line (QED Compton process).
In the phase space considered here,
the Cabbibo-Parisi process, diagram~(b), dominates~(c) due to the pole
of the Compton scattering cross section
for backward scattered photons in the $\gamma e$ centre of mass system.
A similar argument shows that the Drell-Yan process~(d) dominates~(e).
Due to the negative interference of the diagrams~(a) and~(b) in the low mass region ($M_{\mu\mu} <10$~GeV),  
the expected cross section in the analysed
phase space is about 5~$\%$ lower than that calculated when only the contribution of the
two-photon process is considered.

The program GRAPE~\cite{Abe:2001cv} is used to compute
the electroweak theory
predictions. It calculates the cross section according to the diagrams~(a)-(e)
of figure~\ref{fig:feynman:elweak} utilising the GRACE \cite{Ishikawa:1993qr} program.
The program also includes contributions due to $Z^0$ boson exchange.
These contributions become important in cases where
the $Z^0$ radiated from external lines in diagrams~(b)-(e) is close to mass shell.
The $u$-pole contribution to the Drell-Yan-process (diagram d) is neglected.
Its influence was estimated in \cite{Arteaga-Romero:1991wn} and was found to be negligible.
Three different approaches are used to describe the proton structure in different phase space regions for the reaction $ep\rightarrow e\mu\mu X$.  
In the  elastic  region ($X=p$) the hadronic structure is parametrised by the electromagnetic
form factors, which depend on $Q_p^2$, the negative four momentum
transfer squared between the incoming and outgoing proton. 
In the quasi-elastic region ($m_p+m_\pi \leq m_X\leq 5$~GeV) and the soft inelastic 
region  ($m_X> 5$~GeV and $Q_p^2< 1$ GeV$^2$), the calculation is based on parametrisations of the proton
structure functions,
which are given in~\cite{Brasse:1976bf} for the nucleon resonance region ($m_{X}<2$~GeV) and in
\cite{Abramowicz:1997ms} above the resonance region.
In the deep inelastic region ($m_X> 5$~GeV and $Q_p^2\geq 1$ GeV$^2$)  the cross sections for
$eq\rightarrow e\mu\mu q$ are convoluted with the parton density functions of the proton~\cite{Lai:1999wy}.
The GRAPE program is interfaced to 
PYTHIA~\cite{Sjostrand:2001yu} and SOPHIA~\cite{Mucke:2000yb}
for a complete simulation of the inelastic muon pair production processes.
GRAPE allows for QED initial state radiation
by  adapting the cross section calculation in~\cite{Fujimoto:tb}.
Final state radiation is calculated with the parton shower method 
implemented in PYTHIA.

Vector meson production with subsequent decay into muons is another source
of muon pairs in $ep$ scattering. Due to the mass cut ($M_{\mu\mu}>5$~GeV) only the
production of $\Upsilon$ mesons needs to be considered.
The cross section is calculated using the Monte Carlo generator DIFFVM~\cite{List:1993}.
Events with two muons also arise from the decay of tau-leptons produced in two-photon collisions $\gamma \gamma \rightarrow \tau \tau$ and from semi-leptonic decays in open heavy quark production ($Q\bar{Q}$, i.e. $c\bar{c}$ and $b\bar{b}$).
These reactions are simulated  with GRAPE and AROMA~\cite{Ingelman:1997mv}, respectively.
The leading order cross section for $b$-production calculated with AROMA is normalised to the
H1 measurements presented in~\cite{Adloff:1999nr}.

\section{Data Analysis}
\label{sec:analysis}

\subsection{Experimental Conditions}
\label{sec:detector}

The H1 Detector~\cite{Abt:1997hi} contains a central tracking detector  
(full acceptance over the range \linebreak $25^{\circ}  < \theta < 155^{\circ}$)
and a forward\footnote{The forward direction and the positive $z$-axis are given by the proton beam direction. 
Polar angles $\theta$ are defined with respect to the positive $z$-axis. The pseudorapidity is given by $\eta= -\log(\tan{\theta/2})$.} tracking detector ($ 7^{\circ}< \theta < 25^{\circ} $),
which are surrounded by a liquid argon calorimeter ($4^{\circ} < \theta < 154^{\circ}$)
and a lead-scintillating fibre calorimeter\footnote{This device was installed in 1995, replacing a lead-scintillator ``sandwich'' calorimeter~\cite{Abt:1997hi}.} (SpaCal calorimeter~\cite{Appuhn}, $153^{\circ} < \theta < 178^{\circ}$).
The central tracking detector comprises proportional and drift chambers allowing a transverse momentum resolution of $\sigma(P_t)/P_t^2 =0.005$~GeV$^{-1}$. 
These detector components are surrounded by a superconducting magnetic coil with a field strength of 1.15~T. The iron return yoke is
equipped with streamer tubes forming the central muon detector ($4^{\circ} < \theta < 171^{\circ}$).
In the  forward direction, a proton remnant tagger ($0.06^{\circ} < \theta < 0.3^{\circ}$, $z=24$~m) and a forward muon detector
($3^{\circ} < \theta < 17^{\circ}$) are used to separate elastic and inelastic processes.
The trigger for this analysis is based on single muon signatures from the central muon detector, which are combined with signals from the 
central tracking detector.
In events with large hadronic transverse  momenta ($P_{t}^{X} > 12$~GeV) trigger signals from
the liquid argon calorimeter are used in addition.

The procedure to extract cross sections relies on the H1 detector simulation, which is based  
on the GEANT  program~\cite{Geant}.
After simulation, the generated events pass through 
the same reconstruction and analysis chain as the real data.
Trigger and muon identification efficiencies are determined with high statistics data samples for the different 
subdetectors and are incorporated in the simulation.
Acceptances and trigger efficiencies for muon pair production are
then determined from the Monte Carlo simulation and are used to correct the
event yields to obtain cross sections.
The overall trigger efficiency for  di-muon events is about 70~$\%$.
At high hadronic transverse momenta ($P_{t}^{X} > 25$~GeV) this efficiency is above 98~$\%$~\cite{Adloff:1998aw}.

\subsection{Event Selection}
\label{sec:selection}
The event selection and the muon identification are optimised in order to select
an event sample consisting of two well identified muons, isolated from
other objects in the event.
The muon identification~\cite{leissner:2002} is based on measurements from the central tracking detector, the central  muon detector and the liquid argon calorimeter.
Muon candidates are selected from tracks measured in the central tracking detector,
 which are linked to tracks  measured in the muon detector.
Muons which do not reach the muon detector, or enter inefficient regions of the muon detector, 
can be identified by a central track linked to a signature of a minimal ionising particle in the liquid 
argon calorimeter. In about 10~$\%$ of the selected di-muon events, one of the muons is identified only in the
calorimeter.  
The efficiency  for identifying a single muon is typically 75~$\%$ in the kinematic range specified below.
The momentum and angle measurements are obtained from the central tracking detector.

The analysis requires two muons in the phase space  given  by:
\begin{itemize}
\item polar angle region $20^\circ < \theta_{\mu}  < 160^\circ$;
\item transverse momenta $P_t^{\mu_1}>2$~GeV
 and $P_t^{\mu_2}>1.75$~GeV;
\item  invariant mass of the
muon pair $M_{\mu\mu} > 5$~GeV.
\end{itemize}
The polar angular range is matched to the acceptance of the central tracking detector, allowing for a precise
momentum measurement.  The requirement of a  minimum transverse momentum ensures
good muon identification. The analysis is focused on invariant masses above the $J/\psi$-mass.
Low muon pair masses are studied in~\cite{H1:Jpsi,Adloff:2002re}. 

Background from cosmic ray muons is suppressed by requiring that:
\begin{itemize}
\item  the $z$-coordinate of the event vertex is within 40~cm of the nominal interaction point;
\item  the opening angle between the two muons is smaller  than 165$^{\circ}$;
\item  the timing of the event determined in the central tracking detector
coincides with that of the  $ep$ bunch crossing;
\item the timing of the two muon track candidates is consistent with their emergence from a common vertex.
\end{itemize}
The remaining cosmic background contribution is determined to be below 1~$\%$.

An isolation requirement suppresses events with
 muons  from heavy quark decays and events with particles misidentified as muons:
\begin{itemize}
\item the distance  of the muons  to the nearest track or jet\footnote{Jets are considered with minimum transverse momentum of $P_t^{Jet}>5$~GeV, identified with  the  $k_t$-algorithm~\cite{Ellis:tq}.}  in the pseudorapidity-azimuth plane, 
\begin{math} D_{Track,Jet}^{\mu} = \sqrt{\Delta{\eta}^2+{\Delta{\phi}^2}} \end{math}, is required to be greater than \begin{math}  1 \end{math}.
 Since at high transverse momenta the background contributions are small,
 the cut is relaxed to $D_{Track,Jet}^{\mu}>0.5$ for $P_t^{\mu} > 10$~GeV.
\end{itemize}
The remaining contribution of misidentified muons is about 0.5 events, estimated using Monte Carlo simulations of photoproduction and neutral current deep inelastic scattering processes.

The transverse hadronic momentum $P_{t}^{X}$  is measured~\cite{Adloff:2003uh} using 
the liquid argon and SpaCal calorimeters, excluding 
energy deposits of identified muons or electrons.
Electrons are identified in the liquid argon or in the SpaCal calorimeter.

\subsection{Systematic Uncertainties}
\label{sec:systematics}
The following uncertainties on the measured cross sections are taken into account.
\begin{itemize}
\item The uncertainty on the integrated luminosity measurement is $1.5$~$\%$.
\item The uncertainty on the trigger efficiency, obtained from an independently triggered event sample, 
gives a contribution to the systematic uncertainty of 6~$\%$.
\item The uncertainty on the identification efficiency of muons is determined by detailed comparison of data
and simulation efficiencies for a data sample consisting of events with exactly two tracks and at least one identified muon. This leads to 
a contribution to the systematic uncertainty of 6~$\%$.
\item The uncertainty due to the reconstruction efficiency of the central tracking detector for the two muon tracks contributes  with 4~$\%$.
\item The uncertainties on the muon $\theta$ and $\phi$ measurements are 3~mrad and 1~mrad, respectively,
 leading to an effect of up to 1~$\%$ on the cross section.
\item The systematic uncertainty due to biases in the 
transverse momentum measurement for high momentum  tracks ($P_t >20$~GeV) is evaluated
using  the electrons in a neutral current deep inelastic scattering event sample.
A scale  uncertainty is derived from the ratio of the electron energy measured in the calorimeter to its
track momentum measurement.
The  largest effect on the cross section is 7~$\%$ in the highest $P_t$ bin.
\end{itemize}
These uncertainties added in quadrature lead to a total systematic error of 10~$\%$ on the integrated cross section.
The uncertainty on the hadronic energy  scale ($4$~$\%$ for  the liquid argon calorimeter and $7$~$\%$ for the 
SpaCal calorimeter) contributes an additional systematic error to the ${\rm d}\sigma/{\rm d}P_T^X$ determination. 

The uncertainty on the GRAPE calculation is below 1~$\%$ for the elastic process~\cite{abe:private}.
The accuracy of the calculation for the inelastic  process is limited  by the knowledge  of the proton structure.
The uncertainty on the structure function parametrisation (quasi-elastic)
and the parton density function (deep inelastic) cause an uncertainty smaller than 5~$\%$. 
The uncertainty on the  predictions for other sources of muon pair production is estimated to be 30~$\%$ for $ Q\bar{Q} \rightarrow \mu \mu$ and 40~$\%$  for $\Upsilon \rightarrow \mu \mu$.
For the error on the predicted event yields, these theoretical and the experimental uncertainties are added in quadrature with the 
statistical uncertainty on the Monte Carlo calculation.

\section{Results}
\label{sec:results}

\subsection{Inclusive Two Muon Cross Sections}
\label{sec:crosssection:inclusive}

The cross section for the production of events with at least two muons is measured\footnote{This analysis is based on data taken at a centre of mass energy of $\sqrt{s}=319$~GeV. More details can be found in~\cite{leissner:2002}. An analysis of data at $\sqrt{s}=301$~GeV can be found in~\cite{Kaestli}.} and compared with the Standard Model prediction.
In total, 1206 data events with two muons are selected, while $1197 \pm 124$ events are expected according to the Standard Model calculation. No event with more than two muons is observed.
In table~\ref{tab:selection:multimuon:final} the contributions of the different Standard Model processes
are given, where the errors contain the experimental and model uncertainties.
 The electroweak muon pair  production process dominates all other processes,
from which only $ 28.3 \pm 6.7 $ events are expected.
Of the 1206 data events, five events, all with $M_{\mu\mu} < 11$~GeV, have two equally charged muons. 
This observation is in good agreement with the expectation of $4.1 \pm 1.5$ events from the decay of heavy quarks.

\begin{figure}[h]  
\includegraphics[width=7.2cm, viewport=35 0 429 699]{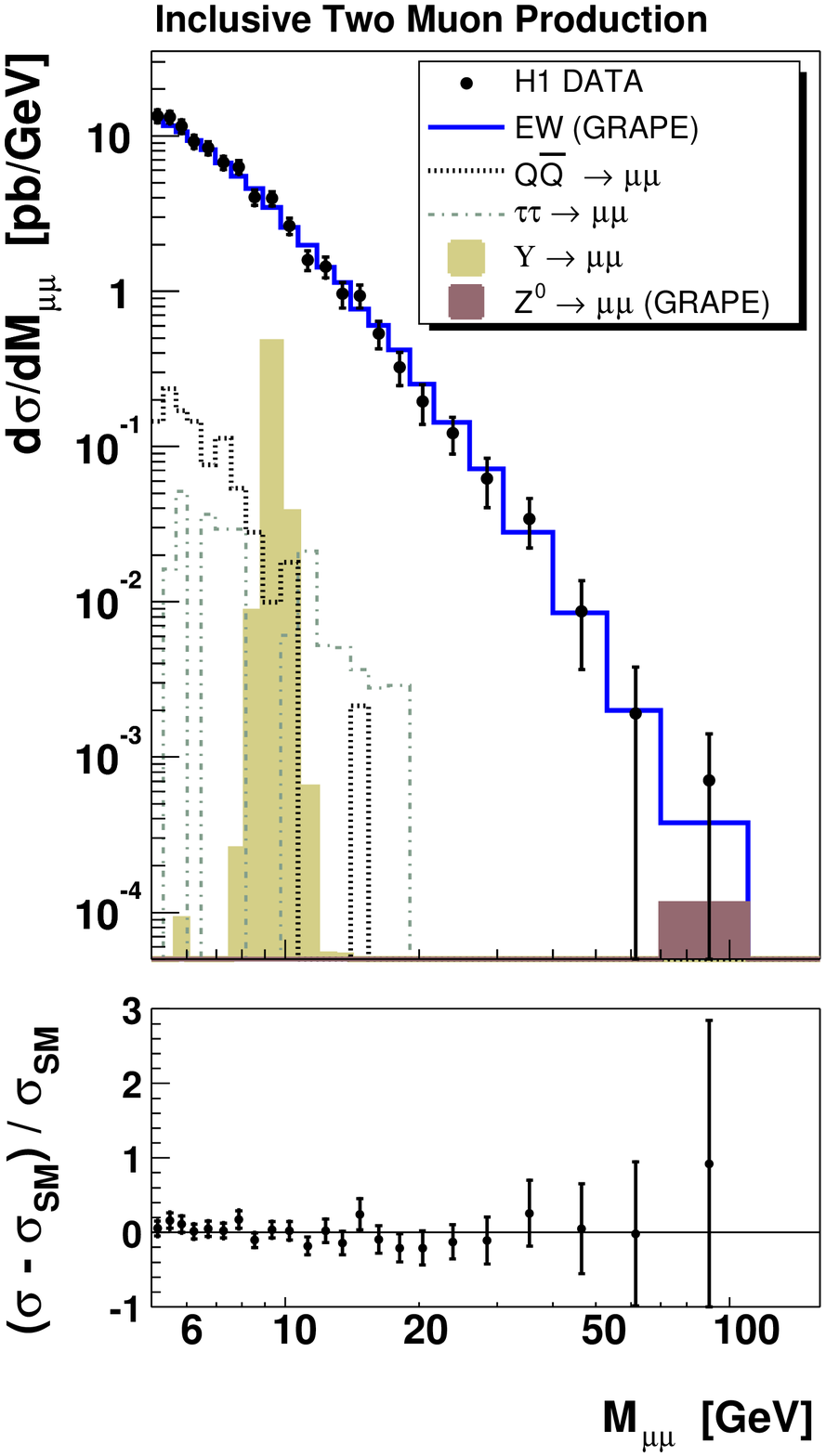}
\hspace{1.1 cm}
\includegraphics[width=7.2cm, viewport=35 0 429 689]{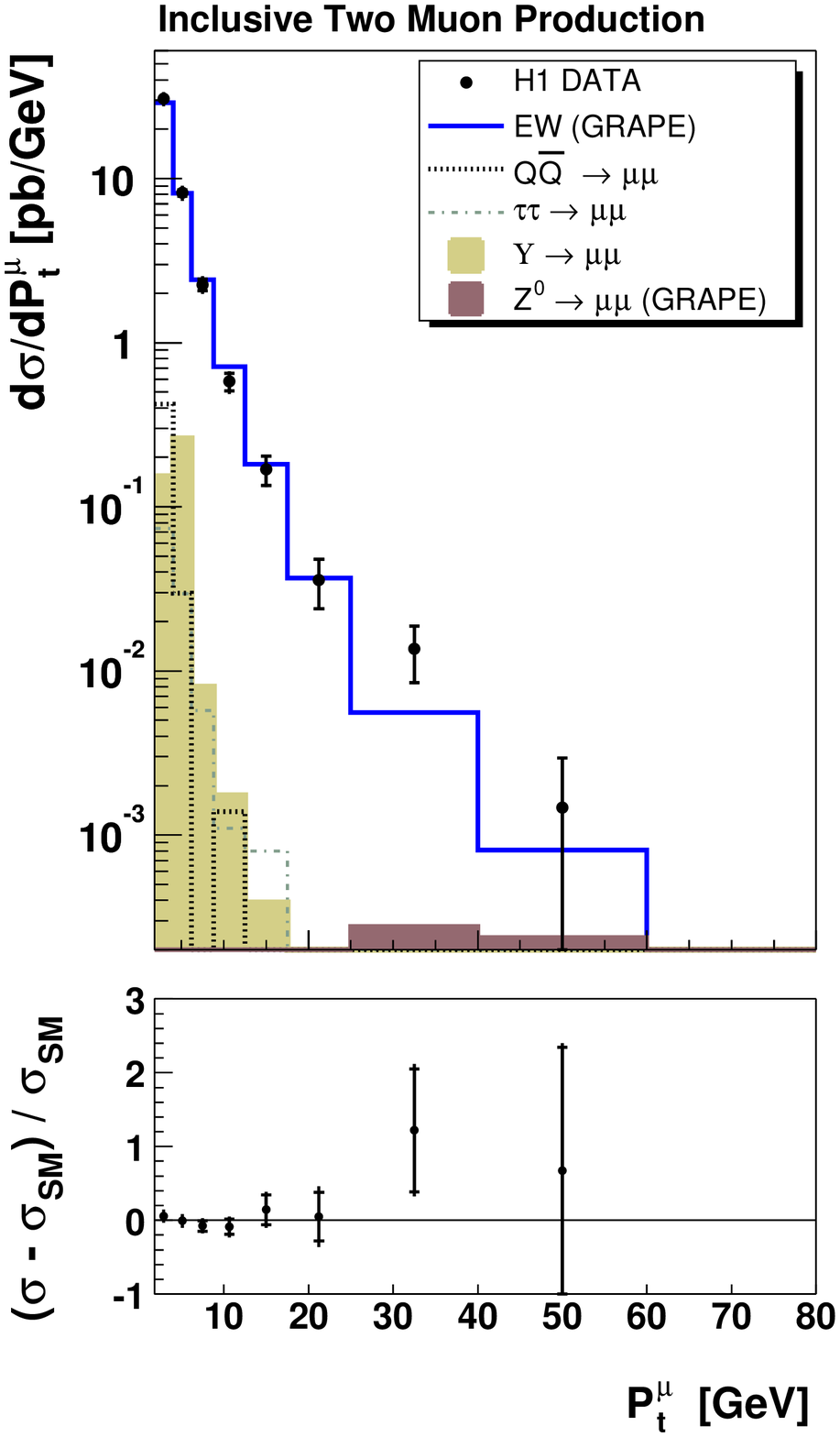}
    \caption{(a) Cross section for the production of two muons in $ep$ interactions as
a function of the di-muon mass $M_{\mu \mu}$. (b) Muon production cross
section as a function of the muon transverse momenta $P_t^\mu$ (two entries per event).
 The data are compared with Standard Model predictions.
See text for the accepted phase space. 
The relative difference between the data and  the sum of all Standard Model contributions is also shown (lower figures). The inner error bars represent the statistical errors, the outer error bars the statistical and systematic errors added in quadrature.}
\label{fig:invariantmass}
\end{figure}

\begin{figure}[h]
  \begin{center}
    \includegraphics[width=14cm]{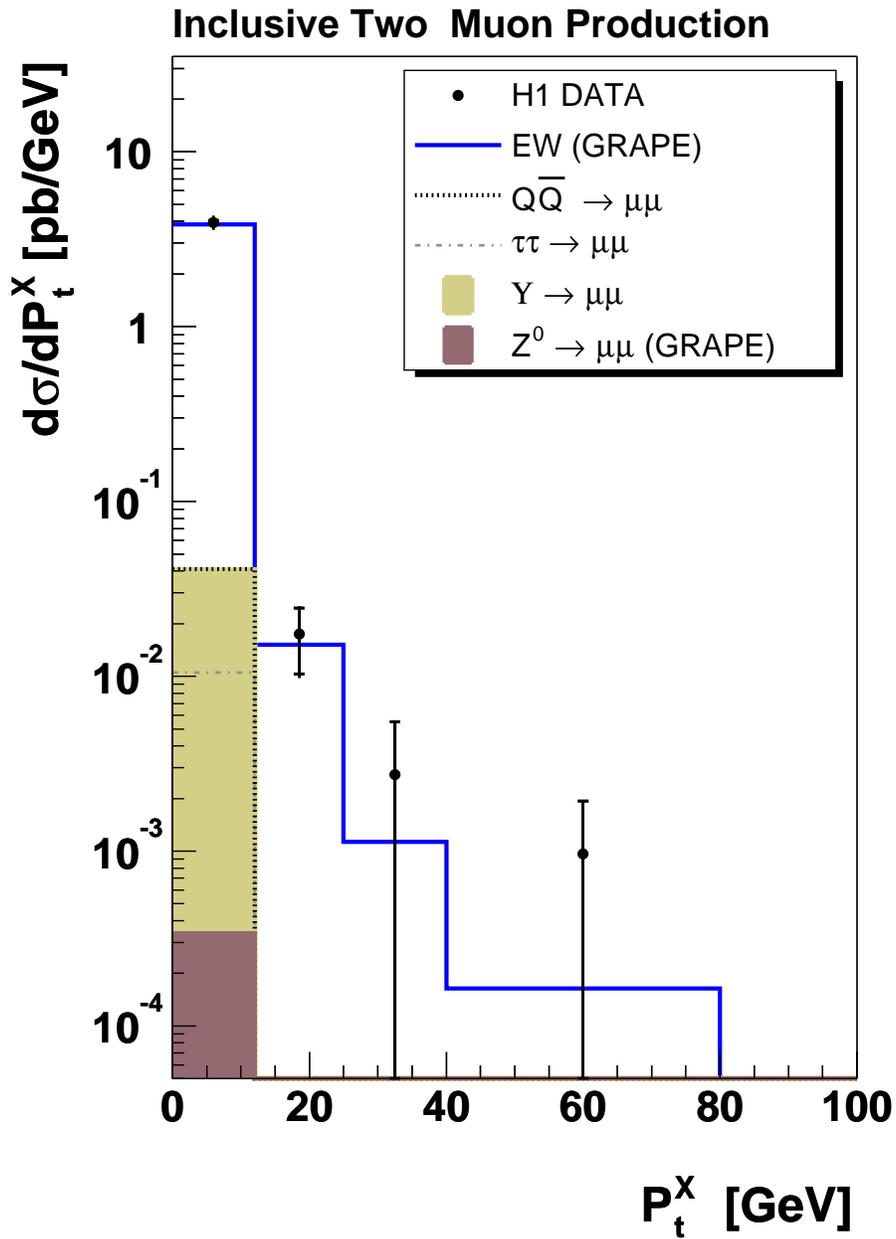}
    \caption{Cross  section for two muon production as a function of the hadronic transverse momentum $P_{t}^{X}$. For further details see  figure \ref{fig:invariantmass}.}
    \label{fig:ptx}
 \end{center}
\end{figure}

The cross section, evaluated in the phase space defined by $M_{\mu\mu}> 5$~GeV, $P_t^{\mu_1} >2$~GeV, $P_t^{\mu_2} >1.75$~GeV and $20^{\circ}< \theta_{\mu} <160^{\circ}$, is presented in figure~\ref{fig:invariantmass}a and table~\ref{tab:xsec:invma} as a function of the di-muon mass $M_{\mu\mu}$.
The cross section falls steeply over more than four decades over the measured mass range, which extends up to 100~GeV.
The shaded  histograms show  the  expected contributions from
the $\Upsilon$  and $Z^0$ resonances, where the latter is also included in the electroweak GRAPE prediction.
At small masses minor contributions from open heavy flavour quark production,
which are strongly suppressed due to the isolation requirement, and  tau-decays are expected.
The muon production cross section as a function of the transverse momenta of the two muons 
is presented  in figure~\ref{fig:invariantmass}b and table~\ref{tab:xsec:pt}.
Both measured cross sections are in good agreement with the Standard Model expectations.
The differential cross section as a function of the  hadronic transverse momentum $P_t^X$ 
is also well described, as shown in figure~\ref{fig:ptx} and table~\ref{tab:xsec:ptx}.

The integrated 
cross section for electroweak muon pair production, $\sigma_{\mu \mu}^{EW}$, 
is obtained by subtracting the expected contributions  
from $\Upsilon$, $Q\bar{Q}$ and $\tau\tau$ decays.
The result is:
\begin{center}
\begin{math}
\sigma_{\mu \mu}^{EW} = (46.4 \pm 1.3  \pm  4.5) \mbox{~pb}. 
\end{math}
\end{center}
The first error gives  the statistical  uncertainty  and  the second  the systematic uncertainty.
The measurement is in good agreement with  the GRAPE  prediction of $(46.1 \pm 1.4)$~pb. 

\subsection{Elastic and Inelastic Muon Pair Production}
\label{sec:crosssection:sep}
Elastic ($ep \rightarrow e\mu\mu p$) and inelastic ($ep \rightarrow e\mu\mu X$)  muon pair production processes
are distinguished by tagging hadronic activity.
An event is assigned as inelastic if activity is detected in the proton remnant tagger, the forward muon detector,
or in the forward region of the liquid argon calorimeter ($\theta < 10^{\circ}$)~\cite{Adloff:2002re}.   
Events containing tracks in the central or forward tracking detectors not associated to the muons or an identified electron 
are also considered as inelastic.   
A total of 631 data events are classified  as elastic  and 575 as inelastic.
This is consistent with the Standard  Model expectation, where $611 \pm 87$ elastic events 
and $586\pm 96$ inelastic events are predicted.
The Monte Carlo simulation shows that
92~$\%$ of generated inelastic events cause activity in the forward detectors and  93~$\%$ of generated 
elastic events remain untagged.
The systematic uncertainty on the separation between elastic and inelastic pair production
takes into  account the tagging efficiencies of the forward detectors.
This leads to an additional uncertainty of 10~$\%$ on the elastic and of 12~$\%$  on the inelastic cross sections.

\begin{figure}[h]
    \includegraphics[width=7.9cm,viewport=15 10 480 700, clip]{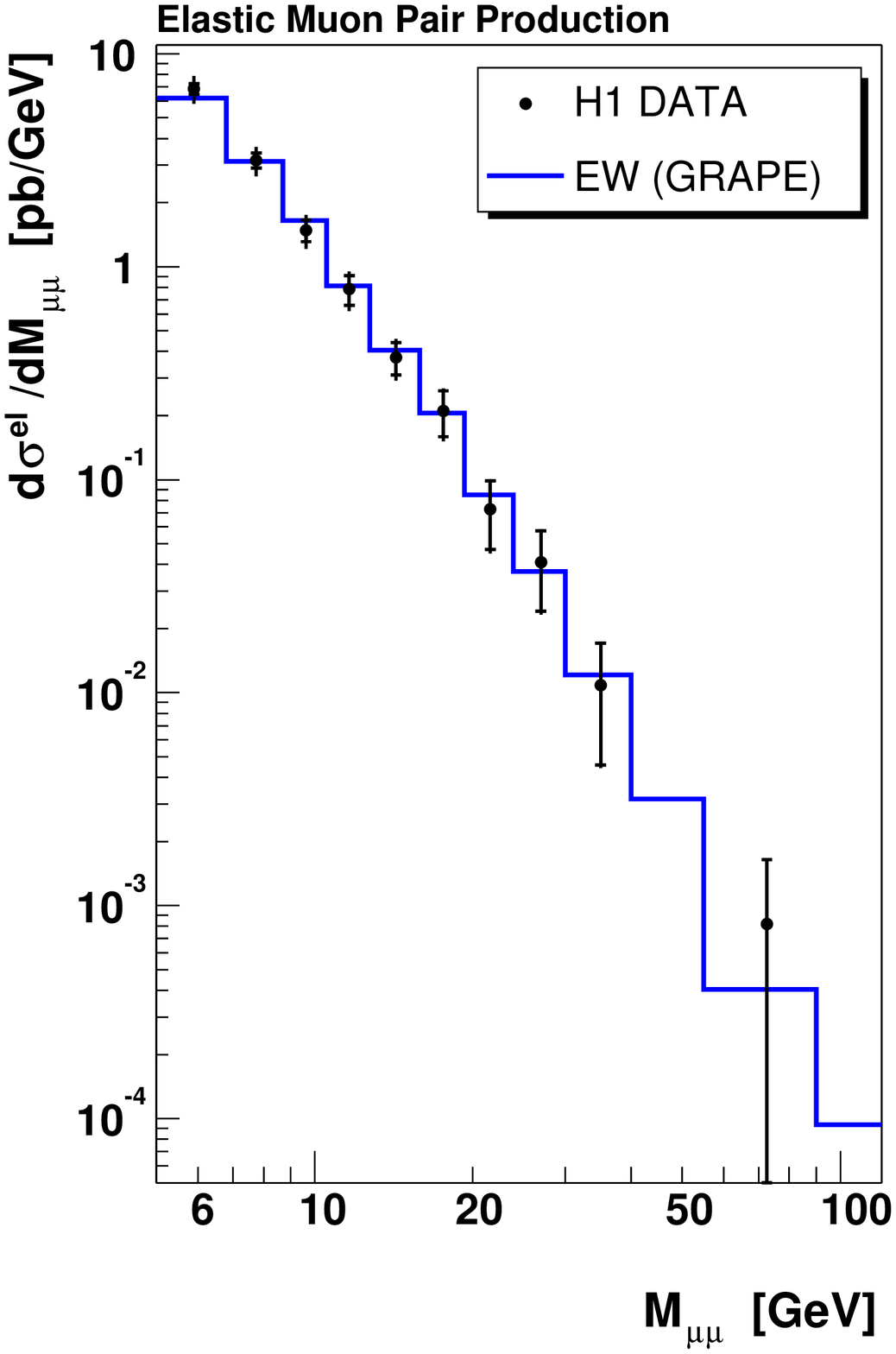}
\hspace{0.5 cm}
    \includegraphics[width=7.9cm,viewport=15 10 480 700, clip]{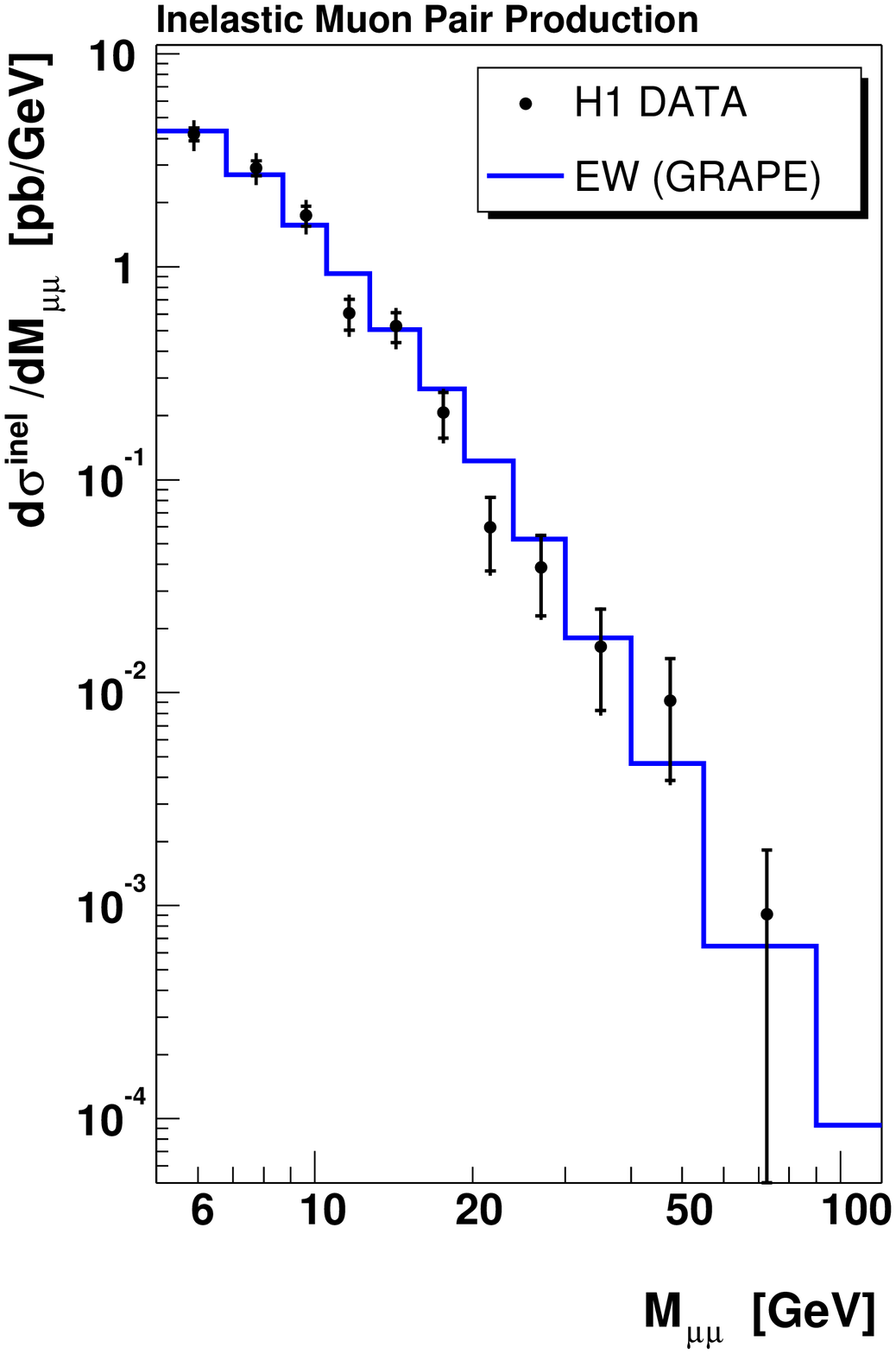}  
 \caption{Cross  section for electroweak (see text) muon pair production as a function of the invariant mass
    $M_{\mu\mu}$ for elastically  produced muon pairs (a) and  for
    inelastically produced muon pairs (b) compared with the electroweak (EW) prediction using the GRAPE generator.
    For the accepted  phase space see text.
    The inner error bars represent the statistical errors, the outer  error bars the statistical
    and systematic errors added in quadrature.}
 \label{fig:ewcrosssectionsela}
 \label{fig:ewcrosssectionsine}

\end{figure}

Figure~\ref{fig:ewcrosssectionsela} and table~\ref{tab:xsec:sep} show the cross sections for  
elastic and inelastic muon pair production after subtraction of $\Upsilon$, $Q\bar{Q}$ and $\tau\tau$ contributions. 
The two spectra are similar and are well described by the electroweak predictions.
Elastic muon production contributes somewhat more in the low mass range 
and inelastic muon production has a slightly harder spectrum. 
This  is expected, as in elastic processes the electromagnetic form factors
of the proton lead to a softer photon spectrum than that produced by radiation from point-like particles
(inelastic  process).

In the analysed phase space an integrated cross section for elastic di-muon production of
\begin{center}
\begin{math}
\sigma_{\mu \mu}^{el} = (25.3  \pm 1.0 \pm  3.5) \mbox{~pb}  
\end{math}
\end{center}
and for inelastic di-muon production of
\begin{center}
\begin{math}
\sigma_{\mu \mu}^{inel} = (20.9  \pm 0.9 \pm  3.2) \mbox{~pb}  
\end{math}
\end{center}
are measured. 
These measurements are in good agreement  with  the expected  cross sections
 of $(24.6 \pm 0.3 )$~pb and $(21.5 \pm 1.1)$~pb, respectively.

\subsection{Multi Lepton Events}
\label{sec:hmass}

In addition to the determination of the inclusive di-muon cross section,
events with two high $P_t$ muons and possible additional leptons, 
either muons or electrons,  have been studied.
In a small fraction of Standard Model electroweak 
di-muon production processes (figure~\ref{fig:feynman:elweak}),
the electron is scattered through a large angle, such that it is visible
in the detector and is not lost in the beam pipe, leading to an observed
$e \mu \mu$ final state. Events with three muons in the final state are
suppressed within the Standard Model and a tri-muon signal would therefore
be of great interest. In order to make use of the highest possible
luminosity, data at $\sqrt{s} = 301 \ {\rm GeV}$ are analysed in addition
to the $\sqrt{s} = 319 \ {\rm GeV}$ sample, resulting in a total luminosity
of $113.7 \ {\rm pb^{-1}}$.

To allow for a comparison with the multi-electron analysis~\cite{H1:electrons},
the following cuts are applied for this study:
\begin{itemize}
\item two muons  in the region  $20^{\circ}<\theta < 150^{\circ}$;
\item transverse momenta $P_t^{\mu_1}>10$~GeV and $P_t^{\mu_2}>5$~GeV.
\end{itemize}
Additional muons must be detected in the central region  of the  detector, $20^{\circ}< \theta_{\mu} < 160^{\circ}$,
with a minimum transverse momentum of $1.75$~GeV. 
Additional electrons are searched for in the polar angle  range  $5^{\circ} < \theta_{e} <  175^{\circ}$
and are required to have a minimum energy of $5$~GeV.
Such $\mu\mu e$ events are triggered with an efficiency of typically $90$~$\%$.

In the examined phase space, 56 di-muon events are found  in the  data, while $54.7 \pm 5.7$ events are  expected.
Among these 56 events, 40 events contain exactly two muons ($\mu \mu$ events), 
compared with $ 39.9 \pm 4.2$ expected. 
In the other 16 events ($\mu \mu e$ events), one additional electron is observed in the liquid argon or the SpaCal calorimeter, compared
with an expectation of $ 14.9 \pm 1.6$ events. 
As expected from the dominant two-photon process, the electron is preferentially found in the backward region of the detector. 
No event with three or more muons or with two muons and more than one electron is observed.

\begin{figure}[h]
  \begin{center}
    \includegraphics[width=16.2 cm]{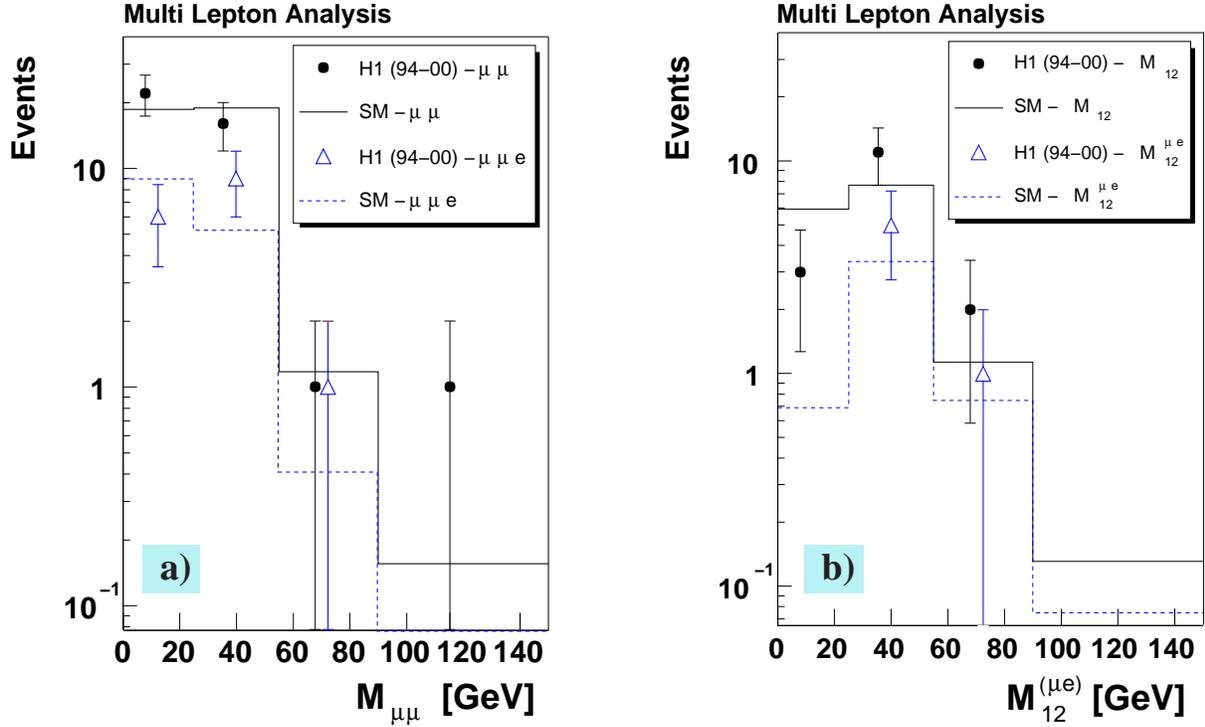}
    \caption{a) Distributions of the invariant mass $M_{\mu\mu}$ for 
           $\mu\mu$ (points) and $\mu\mu e$ (triangles) events, compared with the Standard Model predictions.
           b) $\mu\mu e$ event distributions of the mass 
           $M_{12}$ of the two highest $P_t$ leptons (points), and of the mass  $M_{12}^{\mu e}$ for
           events where  the leptons with the highest $P_t$ are a muon and the electron (triangles).  
           For clarity, the closed data points are shifted slightly to the left.
           The error bars represent the statistical errors. }
    \label{fig:hmmass}
    \label{fig:hm3mass}
 \end{center}
\end{figure}

In figure~\ref{fig:hmmass}a the di-muon mass distributions of events 
classified as $\mu \mu$ events or as $\mu \mu e$ events  are compared with the theoretical expectations.
Both invariant mass distributions are in agreement with the Standard Model calculations.
The distribution in $M_{12}$, the invariant mass of the two leptons
with the largest $P_t$,  is shown for the $\mu \mu e$ sample in figure~\ref{fig:hmmass}b.
This mass combination is selected in order to ease
comparison with the multi-electron analysis~\cite{H1:electrons}, where the scattered electron cannot
be identified uniquely.
For approximately half of these events, the two leptons with the highest $P_t$ are the
electron and  a muon.
For these events, the mass distribution  $M_{12}^{\mu e}$ is also shown
in figure~\ref{fig:hmmass}b.
Both mass distributions are compatible with the Standard Model predictions.

For masses $M_{12} > 100$~GeV ($> 80$ GeV)  one $\mu\mu$ event is found, while $0.08 \pm 0.01$ \mbox{($0.29 \pm 0.03$)} are expected. This inelastic event with two well identified muons 
has a mass of $M_{\mu\mu} = 102\pm11$~GeV and was recorded at $E_p=820$~GeV.
No event classified as $\mu \mu e$ with $M_{12} > 100$~GeV is observed. The prediction is $0.05 \pm 0.01 $. 
These results at high di-lepton masses are in agreement with the Standard 
Model predictions. In view of the present limited statistics, they cannot
be used to draw firm conclusions concerning the high mass excess observed
in the multi-electron analysis~\cite{H1:electrons}.

\section{Conclusion}
Isolated muon pair production is analysed for di-muon invariant masses
above $5 \ {\rm GeV}$. The inclusive, elastic and inelastic cross sections
are measured. In addition, a $\mu \mu e$ event sample is studied. In all
cases, the predictions of the Standard Model are in good agreement with the
observations up to the largest di-lepton masses observed.

\section*{Acknowledgements}

We are grateful to the HERA machine group whose outstanding
efforts have made this experiment possible. 
We thank
the engineers and technicians for their work in constructing and
maintaining the H1 detector, our funding agencies for 
financial support, the
DESY technical staff for continual assistance
and the DESY directorate for support and for the
hospitality which they extend to the non DESY 
members of the collaboration.


\ifpdf
\begin{picture}(1,1)(0,10)
\put(20,174){\colorbox{lightblue}{\large \textbf{ a)}}}
\put(106,174){\colorbox{lightblue}{\large\textbf{ b)}}}
\end{picture}
\else
\begin{picture}(1,1)(0,10)
\put(20,174){\colorbox{lightblue}{\large \textbf{ a)}}}
\put(106,174){\colorbox{lightblue}{\large\textbf{ b)}}}
\end{picture}
\fi

\newpage
\newpage

\begin{picture}(1,1)(0,10)
\put(12,73){\colorbox{lightblue}{\large \textbf{ a)}}}
\put(97,73){\colorbox{lightblue}{\large \textbf{ b)}}}
\end{picture}
\newpage

\begin{picture}(1,1)(0,10)
\put(13,75){\colorbox{lightblue}{\large \textbf{ a)}}}
\put(101,75){\colorbox{lightblue}{\large \textbf{ b)}}}
\end{picture}

\renewcommand{\multirowsetup}{\centering}
\renewcommand{\arraystretch}{1.3}
\renewcommand{\arrayrulewidth}{1.0 pt}
\renewcommand{\arraycolsep}{10 pt}
\begin{table}[htbp]
  \begin{center}
    \begin{tabular}[h]{|c|c!{\vrule width 2pt}c|c|c|c|}
      \hline
\multirow{1}{2 cm}{Data}& \multirow{1}{2 cm}{SM }  & EW \enspace $\mu^{+}\mu^{-}$ & $ \Upsilon \longrightarrow \mu \mu$  & $ \tau \tau \longrightarrow \mu \mu$&$ Q\bar{Q} \longrightarrow \mu \mu$     \\  \hline
  $ 1206   $ &$ 1197 \pm 124 $&$ 1169 \pm 122 $&$ 12.3 \pm 5.1 $ &$   4.5 \pm  0.6 $ &$   11.5 \pm 3.8 $ \\  \hline
    \end{tabular}
    \caption{The number of selected di-muon events compared with the Standard Model prediction (SM).
      The dominant electroweak  contribution (EW) is determined  using  the GRAPE generator. 
      The expectations for other contributions are also  given.}
    \label{tab:selection:multimuon:final}
  \end{center}
\end{table}

\begin{table}[htbp]
  \begin{center}
    \begin{tabular}[h]{|c|c|} \hline
$M_{\mu\mu}$  range   &   ${\rm d}\sigma/dM_{\mu\mu}$\\ 
$[$GeV$]$     &    $[$pb/GeV$]$     \\  \hline
5.0 - 5.3        & $ 13.3 \pm  1.3  \pm 1.3 $ \\
5.3 - 5.7        & $ 12.8 \pm  1.2  \pm 1.2 $ \\
5.7 - 6.0        & $ 12.5 \pm  1.2  \pm 1.2 $ \\
6.0 - 6.5        & $ 9.09 \pm   0.89  \pm 0.86 $ \\
6.5 - 7.0        & $ 8.85 \pm   0.84  \pm 0.84 $ \\
7.0 - 7.6        & $ 7.13 \pm   0.68  \pm 0.68 $ \\
7.6 - 8.2        & $ 6.48 \pm   0.63  \pm 0.62 $ \\
8.2 - 8.9        & $ 4.11 \pm   0.45  \pm 0.39 $ \\
8.9 - 9.8        & $ 4.01 \pm   0.42  \pm 0.38 $ \\
9.8 - 10.7       & $ 2.65 \pm   0.33  \pm 0.25 $ \\
10.7 - 11.8      & $ 1.54 \pm   0.24  \pm 0.15 $ \\
11.8 - 12.9      & $ 1.36 \pm   0.22  \pm 0.13 $ \\
12.9 - 14.1      & $ 0.96 \pm   0.19  \pm 0.09 $ \\
14.1 - 15.4      & $ 0.88 \pm   0.16  \pm 0.08 $ \\
15.4 - 17.1      & $ 0.57 \pm   0.11  \pm 0.05 $ \\
17.1 - 19.1      & $ 0.369 \pm   0.089  \pm 0.035 $ \\
19.1 - 21.6      & $ 0.165 \pm   0.052  \pm 0.016 $ \\
21.6 - 26.0      & $ 0.090 \pm   0.029  \pm 0.009 $ \\
26.0 - 31.0      & $ 0.075 \pm   0.025  \pm 0.007 $ \\
31.0 - 40.0      & $ 0.027 \pm   0.011  \pm 0.003 $ \\
40.0 - 53.0      & $ 0.0093 \pm  0.0054  \pm 0.0009 $ \\
53.0 - 70.0      & $ 0.0019 \pm         0.0019  \pm 0.0002 $ \\
70.0 - 110.0     & $ 0.00075 \pm        0.00075  \pm 0.00009 $ \\ \hline
  \end{tabular}
   \caption{Cross section for the production of two muons in $ep$ interactions 
     as a function of the di-muon mass $M_{\mu\mu}$.
Di-muon events from $\Upsilon$, $\tau$-pair and $Q\bar{Q}$-decays are included in the measurement.
The first  uncertainty  is statistical and the second systematic. }
    \label{tab:xsec:invma}
  \end{center}
\end{table}

\begin{table}[htbp]
  \begin{center}
    \begin{tabular}[h]{|c|c|} \hline 
$P_{T}^{\mu}$  range      & ${\rm d}\sigma/{\rm d}P_{T}^{\mu}$   \\ 
 $[$GeV$]$      &    $[$pb/GeV$]$  \\  \hline
1.8 - 4.0        & $ 30.7 \pm  0.8  \pm 2.9 $ \\
4.0 - 6.2        & $ 8.18 \pm   0.37  \pm 0.78 $ \\
6.2 - 8.8        & $ 2.26 \pm   0.18  \pm 0.22 $ \\
8.8 - 12.5       & $ 0.580 \pm   0.073  \pm 0.055 $ \\
12.5 - 17.5      & $ 0.169 \pm   0.034  \pm 0.016 $ \\
17.5 - 25.0      & $ 0.036 \pm   0.012  \pm 0.003 $ \\
25.0 - 40.0      & $ 0.0136 \pm  0.0052  \pm 0.0014 $ \\
40.0 - 60.0      & $ 0.0015 \pm  0.0015  \pm 0.0002 $ \\ \hline
    \end{tabular}
    \caption{Muon production cross section as a function of  the muon transverse  momenta  $P_t^{\mu}$ (two entries per event). The first  uncertainty  is statistical  and the second systematic.}
    \label{tab:xsec:pt}
  \end{center}
\end{table}

\begin{table}[htbp]
  \begin{center}
    \begin{tabular}[h]{|c|c|}  \hline
 $P_{T}^X$  range  & ${\rm d}\sigma/{\rm d}P_{T}^{X}$   \\ 
 $[$GeV$]$      &    $[$pb/GeV$]$    \\  \hline
0.0 - 12.0       & $ 3.94 \pm   0.11  \pm 0.38 $ \\
12.0 - 25.0      & $ 0.0174 \pm  0.0071  \pm 0.0029 $ \\
25.0 - 40.0      & $ 0.0027 \pm         0.0027  \pm 0.0005 $ \\
40.0 - 80.0      & $ 0.00097 \pm        0.00097  \pm 0.00023 $ \\ \hline
    \end{tabular}
     \caption{Cross section for two muon production as a function  of the hadronic  transverse  momentum $P_t^X$.
The first  uncertainty  is statistical and the second systematic. }
    \label{tab:xsec:ptx}
  \end{center}
\end{table}

\begin{table}[htbp]
  \begin{center}
    \begin{tabular}[h]{|c|c|c|}  \hline
 $M_{\mu\mu}$  range  & ${\rm d}\sigma^{el}/{\rm d}M_{\mu\mu}$  & ${\rm d}\sigma^{inel}/{\rm d}M_{\mu\mu}$ \\ 
 $[$GeV$]$      &    $[$pb/GeV$]$             & $[$pb/GeV$]$     \\ 
\hline 5.0 - 6.8        & $ 6.85 \pm   0.40  \pm 0.95 $ & $ 4.19 \pm   0.29  \pm 0.64 $\\ 
6.8 - 8.7        & $ 3.16 \pm   0.25  \pm 0.44 $       & $ 2.91 \pm   0.24  \pm 0.45 $\\
8.7 - 10.6       & $ 1.48 \pm   0.17  \pm 0.21 $       & $ 1.74 \pm   0.19  \pm 0.27 $\\
10.6 - 12.8      & $ 0.78 \pm   0.13  \pm 0.11 $       & $ 0.60 \pm   0.10  \pm 0.09 $\\
12.8 - 15.9      & $ 0.375 \pm   0.065  \pm 0.052 $       & $ 0.525 \pm   0.084  \pm 0.080 $\\
15.9 - 19.3      & $ 0.210 \pm   0.051  \pm 0.029 $       & $ 0.207 \pm   0.050  \pm 0.032 $\\
19.3 - 23.9      & $ 0.073 \pm   0.026  \pm 0.010 $       & $ 0.060 \pm   0.023  \pm 0.009 $\\
23.9 - 30.0      & $ 0.041 \pm   0.017  \pm 0.006 $       & $ 0.039 \pm   0.016  \pm 0.006 $\\
30.0 - 40.0      & $ 0.0108 \pm  0.0063  \pm 0.0015 $     & $ 0.0165 \pm  0.0082  \pm 0.0025     $\\
40.0 - 55.0     &-----                                  & $ 0.0092 \pm   0.0053  \pm 0.0014 $\\
55.0 - 90.0      & $ 0.00082 \pm         0.00082  \pm 0.00013 $ & $ 0.00091 \pm        0.00091  \pm 0.00017 $\\\hline
\end{tabular}
 \caption{Cross section for electroweak muon pair production as a function  of the invariant mass  $M_{\mu\mu}$  for elastically produced muon pairs (second  column) and inelastically produced muon pairs (third column). 
Muons from $\Upsilon$, $\tau$-pair and $Q\bar{Q}$-decays are considered as background and the  expected event  yields  from these
processes are subtracted from the measured event numbers.
The first  uncertainty is statistical  and the second systematic.}
  \label{tab:xsec:sep}
  \end{center}
\end{table}
\end{document}